\documentclass[prl,aps,epsfig,superscriptaddress]{revtex4}
\usepackage{bm}
\usepackage{amsfonts}
\usepackage[dvips]{graphicx}
\usepackage{mathrsfs}
\usepackage[intlimits]{amsmath}
\usepackage[colorlinks, citecolor=red]{hyperref}

\newtheorem{theorem}{Theorem}

\newtheorem{conjecture}{Conjecture}

\newtheorem{lemma}{Lemma}

\begin{document}

\title{Efficient Representation of Quantum Many-body States with Deep Neural Networks}
\author{Xun Gao$^1$ and Lu-Ming Duan}
\affiliation{Center for Quantum Information, IIIS, Tsinghua University, Beijing 100084,
PR China}
\affiliation{Department of Physics, University of Michigan, Ann Arbor, Michigan 48109, USA}

\begin{abstract}
The challenge of quantum many-body problems comes from the difficulty
to represent large-scale quantum states, which in general requires an
exponentially large number of parameters. Recently, a connection has been made
between quantum many-body states and the neural network representation (\textit{arXiv:1606.02318}). An
important open question is what characterizes the representational power of
deep and shallow neural networks, which is of fundamental interest due to
popularity of the deep learning methods. Here, we give a rigorous proof that a
deep neural network can efficiently represent most physical states, including
those generated by any polynomial size quantum circuits or ground states of
many-body Hamiltonians with polynomial-size gaps, while a shallow network
through a restricted Boltzmann machine cannot efficiently represent those
states unless the polynomial hierarchy in computational complexity theory
collapses.
\end{abstract}
\maketitle

%\email{gaoxungx@gmail.com}

%\textbf{ }

The Hilbert space dimension associated with quantum many-body problems is
exponentially large, which poses a big challenge for solving those problems
even with the most powerful computers. Variational approach is usually the
tool of choice for tackling such difficult problems, which include many
successful examples from simple mean-field approximation to more complicated
methods such as those based on the matrix product states \cite{schollwock2005density,schollwock2011density}, the tensor
network states \cite{verstraete2008matrix,vidal2007entanglement,orus2014practical,carleo2016solving}, the string bond states \cite{schuch2008simulation,sfondrini2010simulating}, and more recently, the neural
network states \cite{carleo2016solving,deng2016exact}. The essence of the variational approach is to find an
efficient representation of the relevant quantum many-body states. Here, by
"efficient" we mean the number of parameters used to characterize those
quantum states increases at most by a polynomial function with the number of
particles (or degrees of freedom) in the system. With an efficient
representation, one can then optimize those variational parameters by
optimization techniques, such as the gradient descent method.

Neural network is a powerful tool to represent complex correlations in
multiple-variable functions and recently finds wide applications in artificial
intelligence through popularity of the deep learning methods \cite{lecun2015deep}. An
interesting connection has been made recently between the variational approach
in quantum many-body problems and the learning method based on neural network
representation \cite{carleo2016solving}. Numerical evidence suggests that the restricted Boltzmann
machine (RBM), a shallow generative neural network, optimized by the
enforcement learning method, provides good solution to several many-body
models \cite{carleo2016solving}. Given this success, an important open question is what
characterizes the representational power and limitation of the RBM for quantum
many-body states.

In this paper, we characterize the representational power and limitation of
the RBM and its extension to deep neural networks, the deep Boltzmann machine
(DBM). We rigorously prove that a DBM can efficiently represent any quantum
states generated by polynomial size quantum circuits or ground states of any
$k$-local Hamiltonians with polynomial-size gaps\textbf{.} Here, "$k$-local"
means that the Hamiltonian has only $k$-body interactions with a finite $k$
(typically small) while the interaction range can be arbitrarily long; and
"polynomial-size gap" means that the energy gap $\Delta$ approaches to zero at
most by $1/$poly$(n)$, where poly$(n)$ denotes a polynomial function of the
particle number $n$. Most physically relevant quantum states are either
generated by many-body dynamics, which can be efficiently simulated through a
polynomial size quantum circuit \cite{lloyd1996universal,poulin2011quantum,berry2015simulating}
, or as ground states of some $k$-local
Hamiltonians. So the DBM can efficiently represent most physical quantum
states. We further prove that those classes of states cannot be efficiently
represented by RBMs unless the polynomial hierarchy, a generalization of the
famous P versus NP\ problem, collapses, which is believed to be highly
unlikely in computer science. While having this limitation, the RBM\ can
indeed represent many highly entangled states, and as examples we give
explicit construction of their representation for arbitrary graph states
\cite{raussendorf2001one}, states with entanglement volume law or for critical
systems \cite{verstraete2006criticality}, and topological toric code states
\cite{kitaev2003fault}.

\begin{figure}[ptb]
\includegraphics[width=0.5\linewidth]{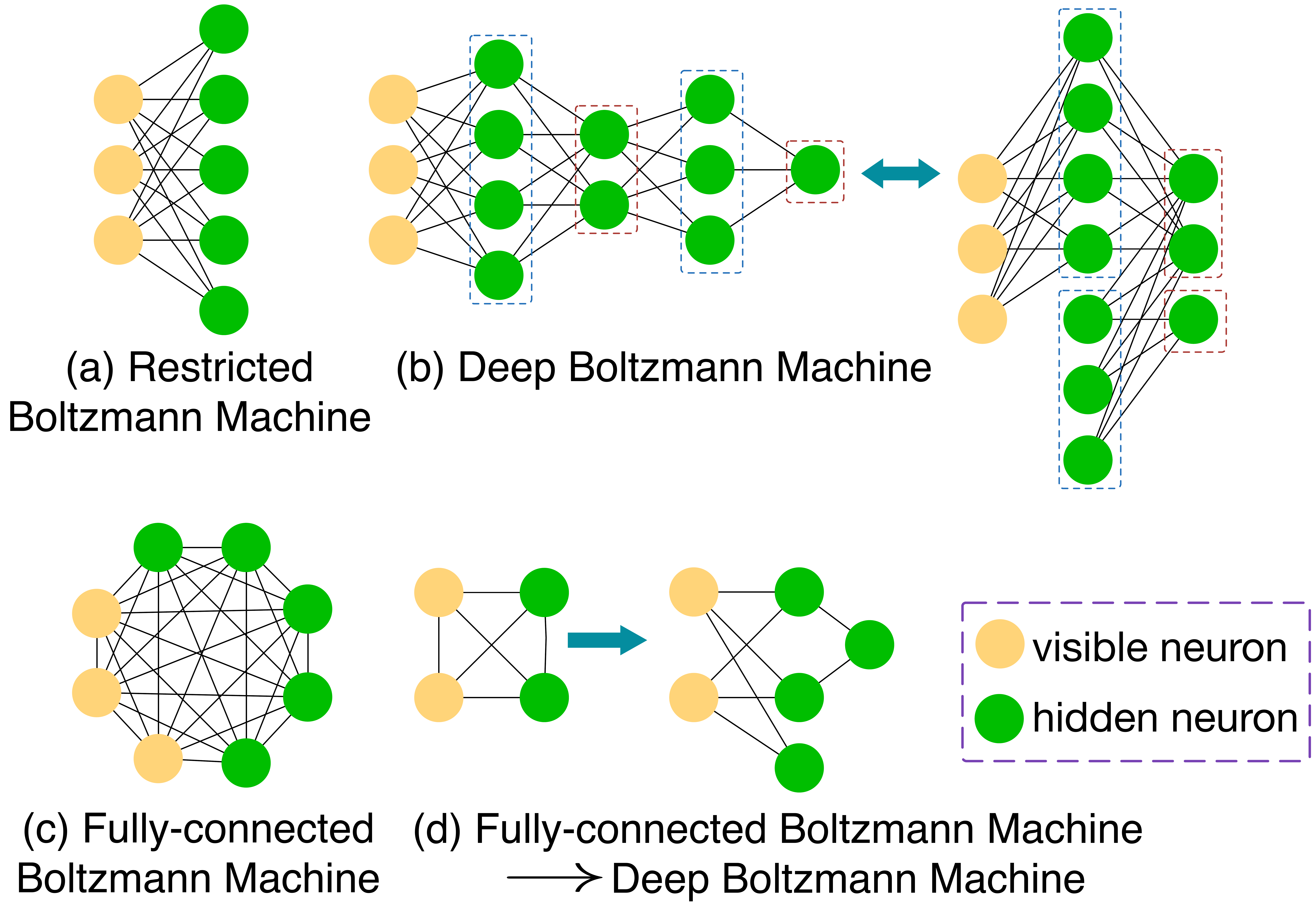}\caption{\textbf{Illustration of
Boltzmann machine neuron networks.} \textbf{a,} Restricted
Boltzmann machine (RBM) which has only one hidden layer and no intra-layer connections.
\textbf{b,} Deep Boltzmann machine (DBM) which has at least two hidden layers and no
intra-layer connections. General DBMs are equivalent to DBMs with two hidden layers after
rearrangement of odd and even layers. \textbf{c,} Fully-connected Boltzmann machine which has
intra-layer connections. \textbf{d,} Reduction of fully-connected Boltzmann machine to DBMs
with two hidden layers.}%
\label{fig:boltzmann}%
\end{figure}

\textbf{Neural network quantum states.} A many-body quantum state of $n$
qubits can be written as $\left\vert \Psi\right\rangle =\sum_{\mathbf{v}}%
\Psi(\mathbf{v)}\left\vert \mathbf{v}\right\rangle $ in the computational
basis with $\mathbf{v}\equiv(v_{1},\cdots,v_{n})$, where the wave function
$\Psi(\mathbf{v)}$ is a general complex function of $n$ binary variables
$v_{i}\in\{0,1\}$. In the neural network representation by a Boltzmann
machine, the wave function $\Psi(\mathbf{v)}$ is expressed as $\Psi
(\mathbf{v})=\sum_{\mathbf{h}}e^{W(\mathbf{v},\mathbf{h})}$, where the weight
$W(\mathbf{v},\mathbf{h})$ is a complex quadratic function of binary variables
$\mathbf{v}$ and $\mathbf{h}\equiv(h_{1},\cdots,h_{m})$ called visible and
hidden neurons, respectively. The number of hidden neurons $m$ is at most
poly$(n)$ for an efficient representation. In the graphic representation shown
in Fig. 1, the neurons $v_{i}$ and $h_{j}$ connected by an edge are correlated
with a nonzero $W_{ij}$ in the weight $W(\mathbf{v},\mathbf{h})=\sum
_{i,j}W_{ij}v_{i}h_{j}$. For the RBM (Fig. 1a), the layer of visible neurons
are connected to one layer of hidden neurons (neurons in the same layer are
not mutually connected). The DBM is similar to the RBM\ but with two or more
layers of hidden neurons (Fig. 1b). Two hidden layers are actually general
enough as one can see in Fig. 1b that odd and even layers can each be combined
into a single layer. A fully-connected Boltzmann machine is shown in Fig. 1c.
In the methods section, we prove that any fully-connected Boltzmann machine can
be efficiently represented by a DBM as illustrated in Fig. 1d.

\begin{figure}[ptb]
\centering
\includegraphics[width=0.5\linewidth]{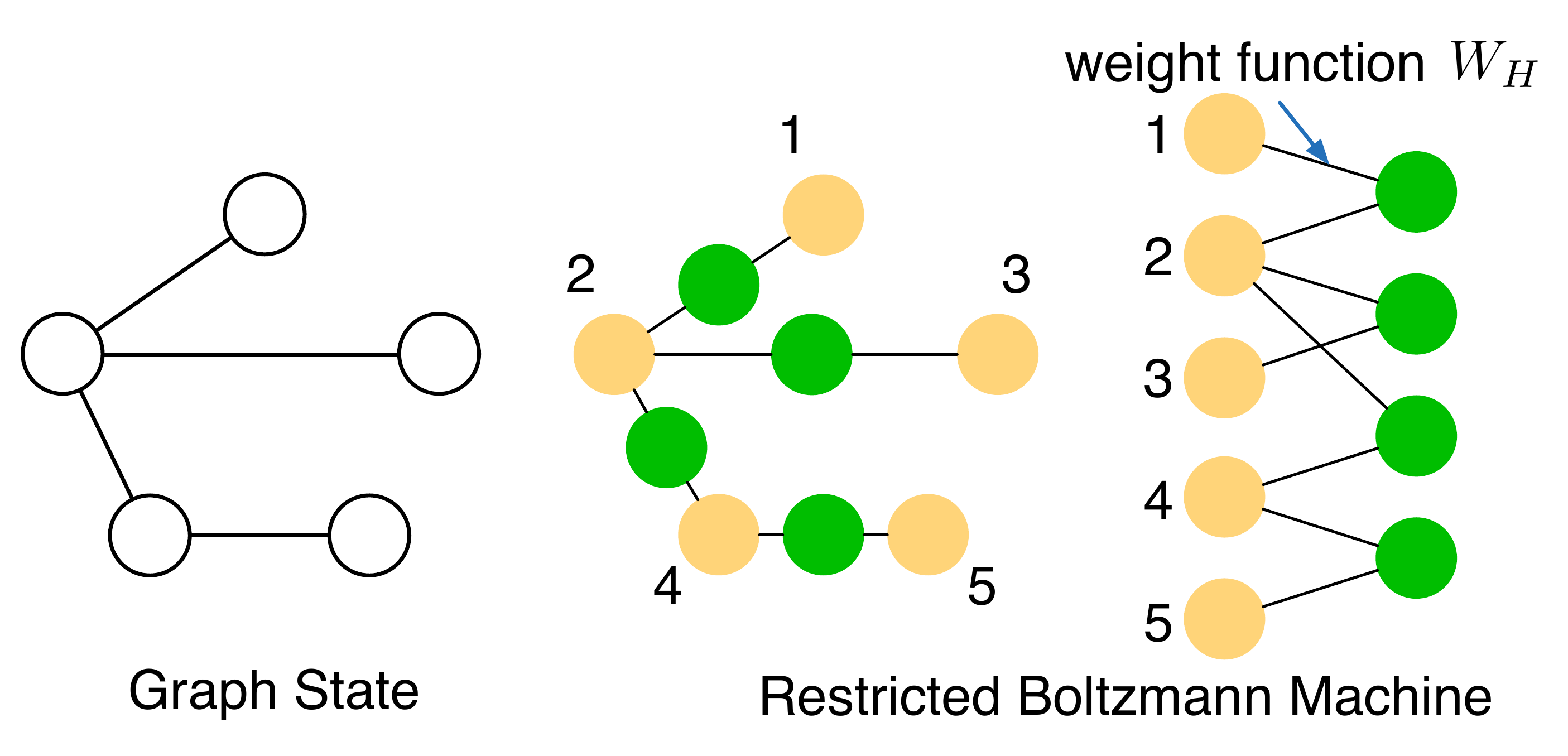}\caption{\textbf{Representation of Graph states
by RBMs.} One hidden neuron with the Hadamard weight function $W_{H}$ (explicit form given in Eq. (1) of the text)
simulates the correlation in the wave function between each pair of connected qubits in any graph states. }%
\label{fig:graph}%
\end{figure}

\textbf{Power and limitation of RBMs.} RBMs can represent a wide
class of many-body entangled states, including wave functions of any graph
states \cite{raussendorf2001one}, toric codes \cite{kitaev2003fault}, and
states violating the entanglement area law or for critical systems
\cite{verstraete2006criticality}. As an example, we give a simple construction
for RBM representation of any graph states and leave the representation of
other categories of states in the Supplementary Information. RBM\ representation for one-dimensional (1D) cluster
states (a special case of graph states) and toric codes have been given
recently in \cite{deng2016exact}. We give a different construction method
which is significantly simpler and more systematic. The wave function of a
graph state takes the form $\Psi(v_{1},\cdots,v_{n})=\prod_{\langle i,j \rangle}%
(-1)^{v_{i}v_{j}}/\sqrt{2}$, where $\langle i,j \rangle$ denotes an edge linking the $i$-th
and $j$-th qubits represented by visible neurons $v_{i},v_{j}$. As shown in
Fig. \ref{fig:graph}, one hidden neuron $h$ and two edges with weight $W_{H}$
realize the correlation function $(-1)^{v_{i}v_{j}}/\sqrt{2}$ between $v_{i}$
and $v_{j}$. This requires to solve the equation $\sum_{h}e^{W_{H}%
(v_{i},h)+W_{H}(v_{j},h)}=(-1)^{v_{i}v_{j}}/\sqrt{2}$, which has a simple
solution
\begin{equation}
W_{H}(x,h)=\frac{\pi}{8}i-\frac{\ln2}{2}-\frac{\pi}{2}ix-\frac{\pi}{4}ih+i\pi
xh\label{1}%
\end{equation}
with $x=v_{i}$ or $v_{j}$.

The RBM\ state has an important property that its wave function $\Psi
(\mathbf{v})$ can be calculated efficiently if a sample of $\mathbf{v}$ is
given (each $v_{i}$ has been assigned a value). Here we prove that this
property leads to limitation of the RBM\ to represent more general quantum
states. With a given sample of $\mathbf{v}$, $\Psi(\mathbf{v})$ can be
factorized as
\begin{equation}
\prod_{j}\left(  \prod_{i:\langle i,j \rangle}e^{W_{ij}(v_{i},0)}+\prod_{i:\langle i,j \rangle}%
e^{W_{ij}(v_{i},1)}\right)  \label{eq:RBM}%
\end{equation}
where $i$ ($j$) runs from $1$ to at most $n$ ($m$), so the total computational
time for $\Psi(\mathbf{v})$ scales as $mn$ for each sample of $\mathbf{v}$.
This means $\Psi(\mathbf{v})$ can be computed by a circuit $C_{n}$ with
polynomial size $\mbox{poly}(n)$ for a given input $\mathbf{v}\in\{0,1\}^{n}$.
If a quantum state has a RBM\ representation (even if its explicit form is
unknown), computing $\Psi(\mathbf{v})$ is characterized by the computational
complexity class $\mathsf{P/poly}$ \cite{arora2009computational}, which
represent problems that can be solved by a polynomial size circuit even if the
circuit cannot be constructed efficiently. The circuit here corresponds to a
RBM representation, with the input given by a specific $\mathbf{v}$ and the
output given by the value of $\Psi(\mathbf{v})$.

We have introduced in Ref. \cite{gao2016quantum} a specific quantum many-body
state, denoted as $\Psi_{\rm GWD}$, for which we proved it is $\mathsf{\#P}$-hard
to calculate its wave function $\Psi_{\rm GWD}(\mathbf{v})$ in the computational
basis $\mathbf{v}$. If this state $\Psi_{\rm GWD}$ has a RBM representation, it
means $\mathsf{\#P}\subset\mathsf{P/poly,}$ an unlikely result in computational
complexity theory as this means the polynomial hierarchy collapses
\cite{karp1982turing,babai1990nondeterministic}. The state $\Psi_{\rm GWD}$ (with
its explicit form given in the Supplementary Information) is just a 2D cluster
state after a layer of translation-invariant single-qubit unitary operations.
This state $\Psi_{\rm GWD}$ is (i) a universal quantum computational state that
can be generated by a polynomial size quantum circuit; (ii) a projected
entangled pair state (PEPS); (iii) the ground state of a gapped $5$-local
Hamiltonian. Combining the results above, we arrive at the following theorem:

\begin{theorem}
RBM cannot efficiently represent universal quantum computational states,
PEPS, and ground states of $k$-local Hamiltonians unless the polynomial
hierarchy collapses.
\end{theorem}

The above argument holds for exact representation of $\Psi(\mathbf{v})$ with
RBM. Similar result holds even if we relax the requirement to have an
approximate representation of $\Psi(\mathbf{v})$ with RBM, i.e., we require
the trace distance between the targeted state and an optimal RBM state bounded
by a small constant. As proved in detail in the Supplementary Information,
under a reasonable complexity conjecture \cite{gao2016quantum}, the
approximate RBM representation of the states listed in Theorem 1 still cannot
be efficient if the polynomial hierarchy does not collapse.

Note that 2D cluster states can be efficiently represented by RBMs. While
after a layer of single-qubit operations which do not change the quantum phase
according to the classification scheme in Ref.\cite{hastings2005quasiadiabatic,chen2010local}, the output state
$\Psi_{\rm GWD}$ cannot be efficiently represented by RBMs any more. So
RBM\ representation is not closed under unitaries that preserve a quantum
phase.

\textbf{Representational power of DBMs}. Now we show with DBMs, i.e., with one
more layer of hidden neurons, all the states listed in Theorem 1, which
include most physical states, can be efficiently represented. For this
purpose, first we introduce a couple of gadgets that will simplify our
construction.

\begin{figure}[ptb]
\centering
\includegraphics[width=1\linewidth]{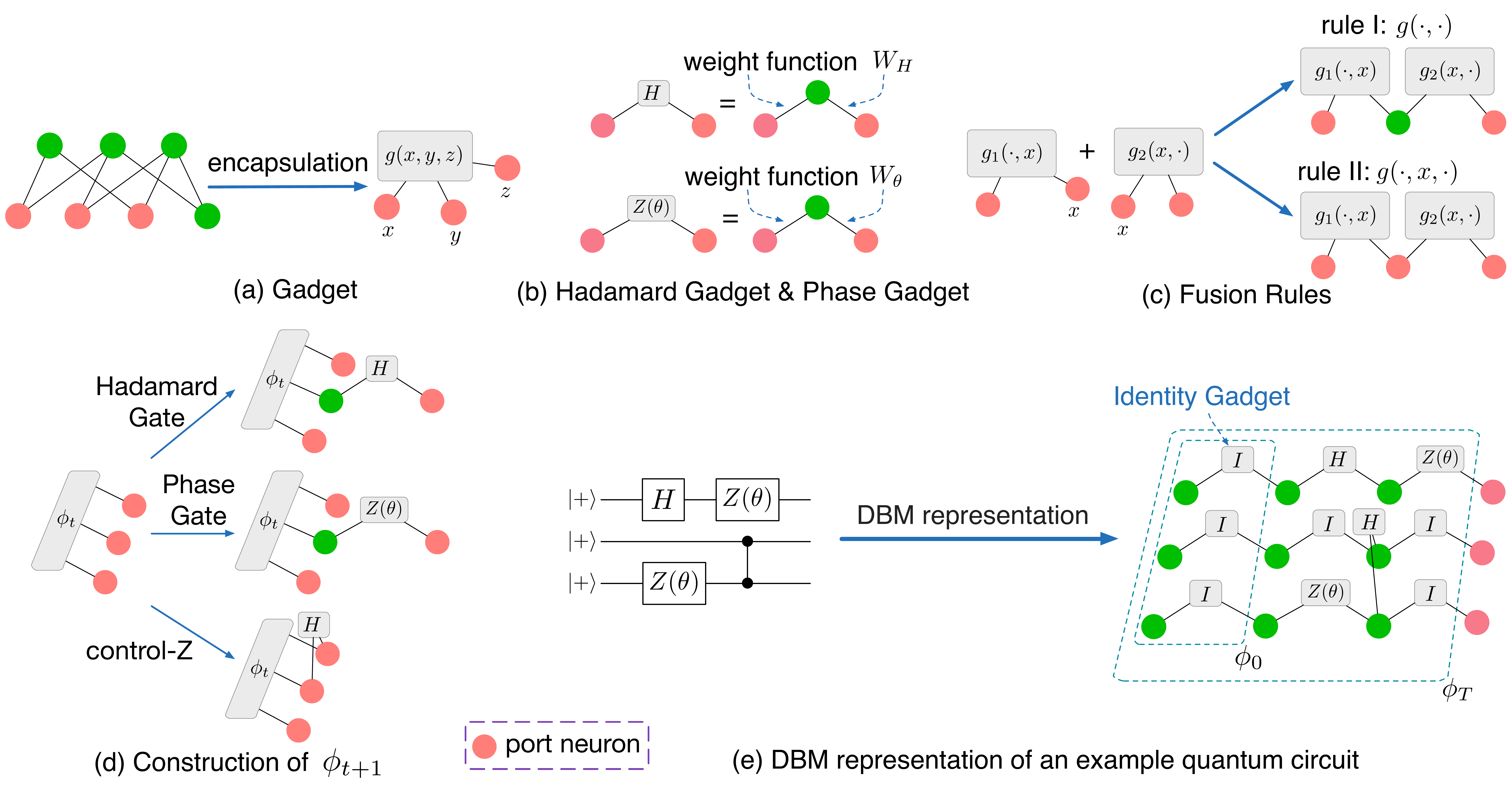}
\caption{\textbf{Representation of universal quantum computational states by DBMs.} \textbf{a,} Gadget is a
complex function of binary variables represented by port neurons, a short-hand notation after
encapsulation of hidden neurons. \textbf{b,} Two elementary gadgets for representation of quantum circuits: Hadamard gadget
with weight $W_{H}$ given by Eq. (1) and phase gadget with weight $W_{\theta}$ given by Eq. (3).
\textbf{c,} Two types of fusion rules for gadgets: rule I and rule II and their neuron network representation.
\textbf{d,} Fusion with phase or Hadamard gadgets with rule I or rule II simulates application of three elementary quantum
gates: the phase gate, the Hadamard gate, and the controlled phase flip gate, which together make universal quantum computation.
The figure illustrates evolution of the wave function from step $t$ to step $t+1$.
\textbf{e,} Representation of an example quantum circuits with elementary gadgets. To represent circuits of depths $T$, we need to
apply $T$ steps of fusions with elementary gadgets, and gadget fusions in the same step can be applied in parallel. The identity gadget is a special phase gadget with $\theta=0$. After the last ($T$) step of computation, port neurons become
visible neurons to represent the index of physical qubits, and we get a DBM representation
of the output state.}%
\label{fig:quantum_circuit}%
\end{figure}

Gadget is a complex function of binary variables after encapsulation of hidden
neurons in a DBM network as shown in Fig. \ref{fig:quantum_circuit}(a), where
the input is represented by port neurons (for connection of different gadgets)
and the output is the value of the function. We use gadgets as basic elements
in a large DBM. As examples, we define Hadamard gadget and phase gadget as
shown in Fig. \ref{fig:quantum_circuit}(b), which will play the role of
elementary gates for construction of DBM representation of quantum circuits.
The weight function $W_{H}$\ is given by Eq. (1) and $W_{\theta}$ is the
solution of the equation $\sum_{h}e^{W_{\theta}(x_{1},h)+W_{\theta}(x_{2}%
,h)}=e^{i\theta x_{1}}\delta_{x_{1}x_{2}}$, which may take the form
\begin{equation}
W_{\theta}(x,h)=-\frac{\ln2}{2}+\frac{\theta}{2}ix+i\pi xh.
\end{equation}
We can combine two gadgets $g_{1},g_{2}$ into one gadget $g$ by two types of
fusion rules shown in Fig. \ref{fig:quantum_circuit}(c):
\begin{align}
\mbox{rule I: } &  g(\cdot,\cdot)=\sum_{x}g_{1}(\cdot,x)g_{2}(x,\cdot
),\label{eq:ruleI}\\
\mbox{rule II: } &  g(\cdot,x,\cdot)=g_{1}(\cdot,x)g_{2}(x,\cdot
),\label{eq:ruleII}%
\end{align}
where rule I simulates matrix multiplication.

With these tools, now we construct efficient DBM representation of any quantum
states generated by a polynomial size circuit. The Hadamard gadget and phase
gadget as shown in Fig. \ref{fig:quantum_circuit}(b) are used to construct
three elementary quantum gates: Hadamard gate $H$, phase gate $Z(\theta)$ with
an arbitraray phase $\theta$, and controlled phase flip gate $CZ$, which
together are universal for quantum computation
\cite{barenco1995elementary,nielsen2010quantum}. The initial state of the
circuit is taken as $\left(  |0\rangle+|1\rangle\right)  ^{\otimes n}$, an
equal superposition of computational basis states, which is represented by the
wave function $\phi_{0}(x_{1},\cdots,x_{n})=1$, the identity gadget.  Denote
the wave function after applying $t$-layer of elementary gates as $\phi_{t}%
(x_{1},\cdots,x_{n})$. As shown in Fig. \ref{fig:quantum_circuit}(d), using
rule I (corresponding to matrix multiplication), Hadamard gadget and phase
gadget simulate gates $H$ and $Z(\theta)$. Using rule II with Hadamard gadget,
we have
\begin{equation}
\phi_{t+1}(\cdots x_{i},x_{i+1}\cdots)=(-1)^{x_{i}x_{i+1}}\phi_{t}(\cdots
x_{i},x_{i+1},\cdots)/\sqrt{2},
\end{equation}
which simulates the $CZ$ gate except for the unimportant normalization factor
$1/\sqrt{2}$. The above procedure can be paralleled as illustrated in Fig.
\ref{fig:quantum_circuit}(e), which shows the DBM\ representation of an
example circuit. For a quantum circuit of depth $T$, we apply $T$\ steps of
fusion rules, and each step needs $O(n)$ neurons. So the DBM representation of
the output state of the quantum circuit takes $O(nT)$ neurons. This
DBM\ representation is sparse, meaning that each neuron has a constant
coordination number (number of connected edges) that does not increase with
the size of neuron network. We therefor have the following theorem:

\begin{theorem}
\label{thm:quantum_circuit} Any quantum state of $n$ qubits generated by a
quantum circuit of depth $T$ can be represented exactly by a sparse DBM with
$O(nT)$ neurons.
\end{theorem}

\begin{figure}[ptb]
\centering
\includegraphics[width=0.5\linewidth]{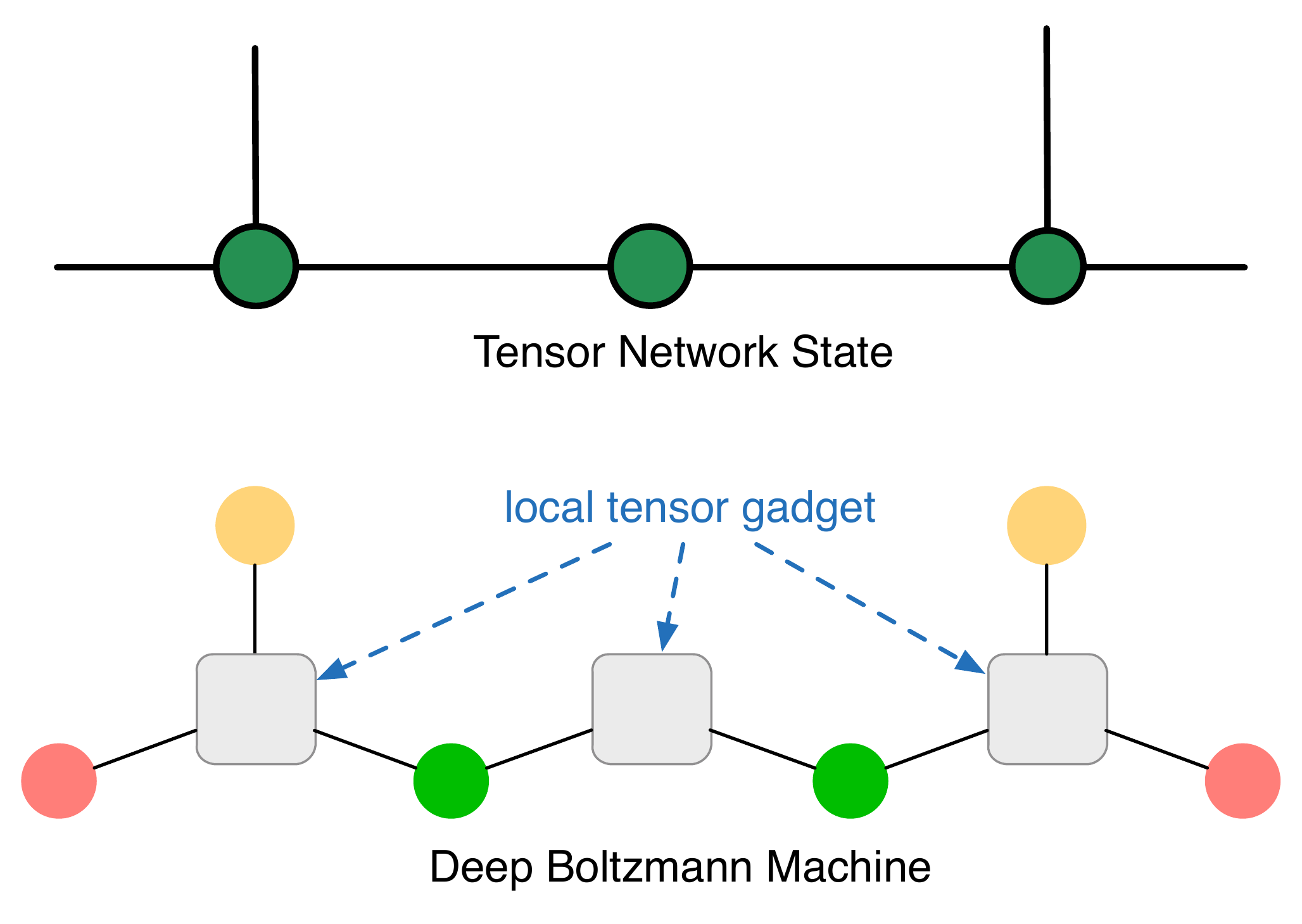}\caption{\textbf{Representation of
tensor network states with DBMs.} Visible (hidden) neurons play the role of physical (bond) indices,
respectively. Port neuron represents either the bond index for the next step of tensor contraction or the physical
index if there is no further contraction. The grey box stands for the local
tensor gadget $A_{x_{1}\cdots x_{c}}$ which can be efficiently represented with a DBM.}%
\label{fig:tensor_network}%
\end{figure}

Using the above theorem, we now construct efficient DBM representation of any
tensor network states, which include the PEPS and the MERA states as special
cases \cite{verstraete2004renormalization,vidal2007entanglement,bridgeman2016hand}. Suppose the local tensor is $A_{b_{1}\cdots b_{d}p}$, which has one
(or zero) physical index $p$ and $d$ bond indices $b_{1},\cdots,b_{d}$, each
ranging from $1$ to the bond dimension $D$. Without loss of generality, we
assume $p$ is binary and $D=2^{k}$ for some integer $k$ and write the local
tensor as a function $A_{x_{1}\cdots x_{c}}$, where each $x_{i}$ is a binary
variable and $c=kd+1$. The state of $|A\rangle=\sum_{x_{1},\cdots,x_{c}%
}A_{x_{1}\cdots x_{c}}|x_{1},\cdots,x_{c}\rangle$ can be generated by a
quantum circuit with the number of elementary gates on the order of
$O(2^{2(kd+1)})=O(D^{2d})$ \cite{nielsen2010quantum}, which is square of the
Hilbert space dimension of $\mbox{span}(|x_{1},\cdots,x_{c}\rangle)$. Using
Theorem 2, the state $|A\rangle$ can be represented by a DBM\ with $O(D^{2d})$
neurons, and the resultant representation is called the local tensor gadget.
We use fusion rule I to link two local tensor gadgets to simulate contraction
of bond index and put physical index in the visible layer, as shown in Fig.
\ref{fig:tensor_network}. We thus have the following theorem:

\begin{theorem}
\label{thm:tensor_network} Tensor network state with bond dimension $D$,
maximum coordination number $d$, and $n$ local tensors, can be represented
efficiently by a sparse DBM with $O(nD^{2d})$ neurons.
\end{theorem}

Tensor network state can represent ground state of a Hamiltonian by simulating
imaginary time evolution through Trotter decomposition
\cite{schuch2007computational,vidal2003efficient,orus2008infinite}. Recently,
quantum simulation based on truncated Taylor series has been proposed
\cite{berry2015simulating} which has exponential improvement on precision
compared to traditional methods based on Trotter decomposition. Inspired by
this idea, we construct tensor network simulation for imaginary time evolution
of any $k$-local Hamiltonian based on truncated Taylor series. Compared to
previous method \cite{schuch2007computational}, our construction offers
exponential improvement on precision. The detailed proof is included in the
Supplementary Information.   Combining with theorem \ref{thm:tensor_network},
we then construct an efficient DBM representation of ground state of any
$k$-local Hamiltonian, described by the following theorem:

\begin{theorem}
The ground state of any $k$-local Hamiltonian can be represented by a sparse
DBM with neuron number
\begin{equation}
O\left(  \frac{1}{\Delta}\left(  n+\log\frac{1}{\epsilon}\right)
m^{2}\right),
\end{equation}
where $n$ is the particle number, $m$ is number of interaction terms in the
Hamiltonian, $\Delta$ is the energy gap, and $\epsilon$ is the
representational error in terms of trace distance.
\end{theorem}
This representation is efficient as long as the energy gap $\Delta$ vanishes with increase of $n$  no faster than $1/poly(n)$,
which is typically true for physical Hamiltonians (even if they are gapless in the thermodynamic limit).

\textbf{Discussion.} With popularity and success of the deep learning methods, a question
often raised is why depth of a neural network is so important \cite{lecun2015deep}. Our proof of the exponential gap in
efficiency of using the DBMs and the RBMs to represent quantum many-body states helps to address this fundamentally important question
in a new context of solving problems in the quantum world. We have proven that most physical quantum states, either from quantum
dynamics or as ground states of complicated Hamiltonians, can be efficiently
represented by DBMs. This result is of fundamental interest and may open up an exciting prospect of using deep
learning methods through neural network representation to tackle strongly
correlated many-body problems, a challenging frontier of modern physics.

\section{Methods}

Here we prove that any fully-connected Boltzmann machine (with intra-layer
edges) can be efficiently simulated with DBMs (without intra-layer connections)
as shown in Fig. 1(d). The key point is to simulate the interaction between
two neurons by a gadget $\sum_{h}e^{W_{1}(x_{1},h)+W_{2}(x_{2},h)}%
=e^{W_{0}(x_{1},x_{2})}$. Suppose the interaction term is $Jx_{1}x_{2}$ in
$W_{0}$, we need $W_{1}+W_{2}=a-\ln2+b(x_{1}+x_{2})(2h-1)+c(2h-1)+d(x_{1}+x_{2})$ to simulate the
interaction with the aid of hidden neuron $h$, where the parameters $a,b,c,d$ need
to satisfy the equations $e^{a}\cosh(c)=1$, $e^{a}e^{d}\cosh(b+c)=1$, $e^{a}%
e^{2d}\cosh(2b+c)=e^{J}$. These equations have solutions, one of them is
\begin{equation}%
a=-d=-J/2,b=-c=-i\arccos(e^{J/2}).
\label{eq:FBM}%
\end{equation}

\textbf{Acknowledgements} We thank Ignacio Cirac and Sheng-Tao Wang for helpful discussions.
This work was supported by the Ministry of Education of China and Tsinghua University.
LMD acknowledges in addition support from the AFOSR MURI program.

\textbf{Author Information} All the authors contribute substantially to this work.
Correspondence and requests for materials should be addressed to L.M.D.
(lmduan@umich.edu) or X. G. (gaoxungx@gmail.com).

\section{Supplementary Information}

In this Supplementary Information, we provide details on derivations and
proofs in main text, including, (i) a detailed derivation of the
weight functions $W_{H}$ and $W_{\theta}$; (ii) construction of RBM
representation for toric codes and for states with entanglement volume law or
critical behaviors. (iii) a proof of theorem 1 for limitation of RBM in the
case of approximate representation; (iv) a proof of theorem 4 for DBM
representation of ground states of any $k$-local Hamiltonians.

\subsection{Detailed derivation of the weight functions $W_{H}$ and $W_{\theta}$}

In main text, we give the expression for $W_{H}$ and $W_{\theta}$ which can be
obtained by setting a general form for them as $a+bx+ch+dxh$ and solving the
resultant equations for the parameters $a,b,c,d$. Here, we give the detailed derivation.

For the Hadamard gadget which is used to construct RBM representation for
graph states and simulate $H$ and $CZ$ gates, the equation we need to solve
is
\begin{equation}
\sum_{h=0,1}e^{W_{H}\left(  x_{1,}h\right) + W_{H}\left(  x_{2,}h\right)}
=H_{x_{1}x_{2}}%
\end{equation}
where the correlation
\begin{equation}
H_{x_{1}x_{2}}=\frac{(-1)^{x_{1}x_{2}}}{\sqrt{2}}=\cos\left(  \frac{\pi}%
{4}\left[  2(x_{1}+x_{2})-1\right]  \right)  .
\end{equation}
The last step in Eq.(2)\ is valid since we have $H_{x_{1}x_{2}}=1/\sqrt{2}$,
$1/\sqrt{2}$, $-1/\sqrt{2}$ when $x_{1}+x_{2}=0,1,2$, respectively. Using the
relation
\begin{equation}
\cos X=\frac{e^{iX}+e^{-iX}}{2}=\sum_{h}e^{iX(2h-1)-\ln2}.
\end{equation}
with $X=\pi/4\left[  2(x_{1}+x_{2})-1\right]  $, we get%
\[
W_{H}\left(  x_{i,}h\right)  =i\pi x_{i}h-i\pi\left[  2x_{i}+h\right]
/4+\left(  i\pi/4-\ln2\right)  /2.
\]
for $x_{i,}=x_{1}$ or $x_{2}$.

For the Phase gadget which is used to simulate $Z(\theta)$ gate, we need to
satisfy
\begin{equation}
\sum_{h=0,1}e^{W_{\theta}\left(  x_{1,}h\right) + W_{\theta}\left(  x_{2,}%
h\right) } =Z(\theta)_{x_{1}x_{2}}=\delta_{x_{1}x_{2}}e^{i\theta x_{1}}%
\end{equation}
We simulate $\delta_{x_{1}x_{2}}$ by the following observation
\begin{equation}
\delta_{x_{1}x_{2}}=\frac{1+e^{i\pi(x_{1}+x_{2})}}{2}=\sum_{h}e^{i\pi
(x_{1}+x_{2})h-\ln2}.
\end{equation}
Note that $\delta_{x_{1}x_{2}}e^{i\theta x_{1}}=\delta_{x_{1}x_{2}}%
e^{i\theta(x_{1}+x_{2})/2}$, we have the solution
\begin{equation}
W_{\theta}\left(  x_{i,}h\right)  =i\pi x_{i}h+\left(  i\theta x_{i}%
-\ln2\right)  /2.
\end{equation}

\subsection{RBM representation of many-body entangled states}

\begin{figure}[ptb]
\centering
\includegraphics[width=1\linewidth]{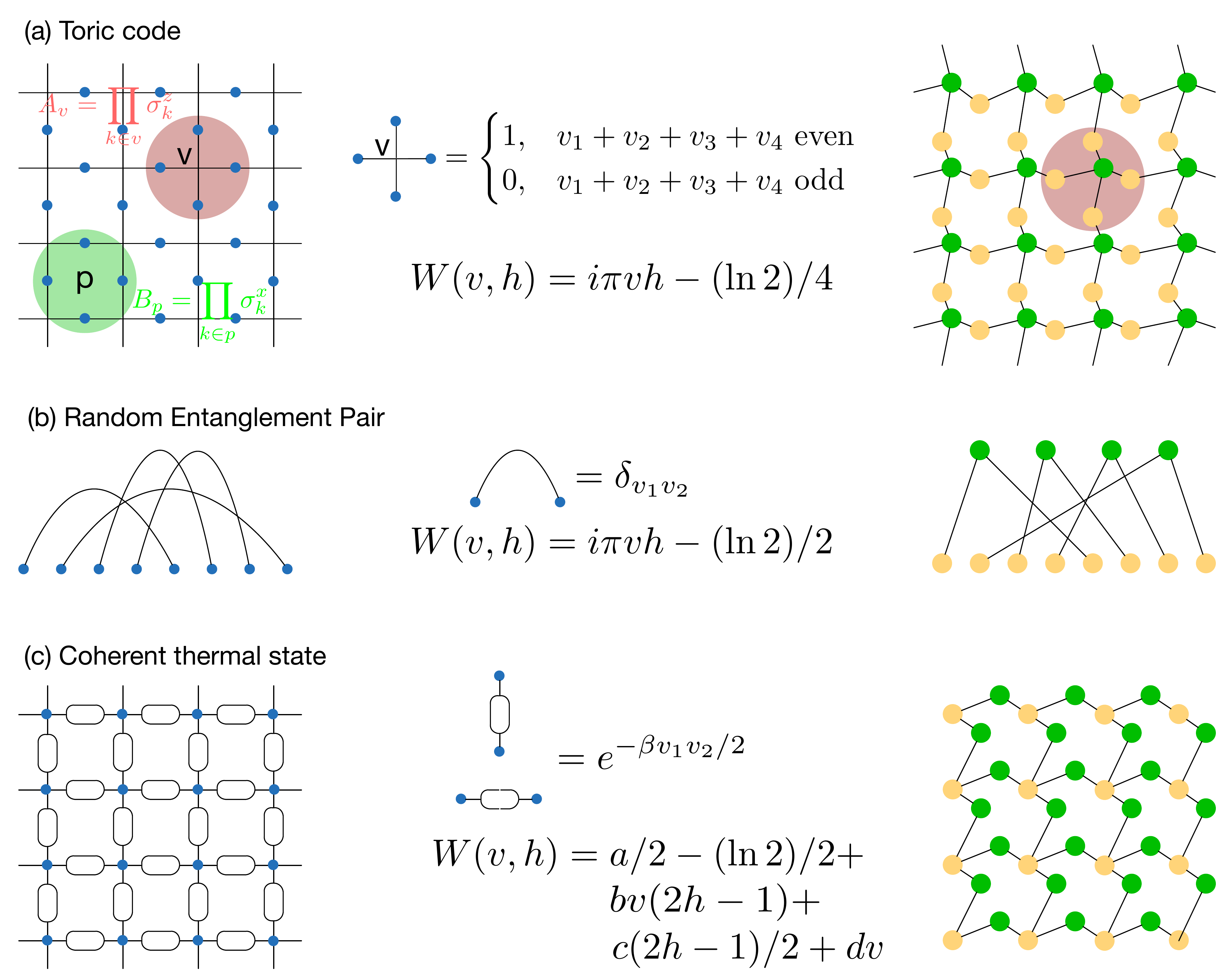}\caption{RBM
representation (shown on the right side) of some many-body entangled states (defined on the left side). 
(a) Toric code is the simplest topologically ordered state
and is used as a quantum error correcting code. The wave function for the toric code is a product of
functions shown in the figure for each vertex $v$. (b) Randomly distributed Entangled pairs
$|00\rangle+|11\rangle$ on a line (or any lattice). The state obeys the entanglement volume law.
The wave function of this state is a simple product of functions shown in the figure for each pair.
(c) Coherent thermal state which represents a critical system when $\beta$ reaches
$\beta_{c}$, the critical point in the corresponding thermal state of
the classical Ising model. The wave function is square root of the corresponding probability in
classical thermal model. For RBM representation, the coefficient $a,b,c,d$ is given by the solution in the 
method section of the main text with $J$ replaced by $-\beta/2$.}%
\label{fig:examples}%
\end{figure}

In this construction, we restrict to simple RBM gadget with only one hidden
neuron connected to $k$ visible neurons and identical weight function $W$ on
each edge. we need to solve the equation
\begin{equation}
\sum_{h}e^{W(v_{1},h)+\cdots+W(v_{k},h)}=g(v_{1},\cdots,v_{k}%
)\label{eq:gadget}%
\end{equation}
for a certain correlation $g(v_{1},\cdots,v_{k})$. Apart from the graph state
example given in the main text, we construct RBM\ representation of three
classes of entangled states: the toric code, which is the simplest state with
topological order and useful for quantum error correction; the randomly
distributed entangled pair state, which satisfies the entanglement volume law
instead of the area law \cite{vitagliano2010volume}; and the coherent thermal state which describes a
critical system~\cite{verstraete2006criticality}. These examples and the corresponding RBM representations
are shown in Fig. \ref{fig:examples}.

For the toric code, the correlation function $g(v_{1},v_{2},v_{3}%
,v_{4})=(v_{1}+v_{2}+v_{3}+v_{4}\mod 2)$ on each vertex and the wave function
is a product of these functions over all vertices \cite{gu2009tensor}. We
have
\begin{equation}
g(v_{1},v_{2},v_{3},v_{4})=\frac{1+e^{i\pi(v_{1}+v_{2}+v_{3}+v_{4})}}{2}%
=\sum_{h=0,1}e^{i\pi(v_{1}+v_{2}+v_{3}+v_{4})h-\ln2},
\end{equation}
so the weight $W(v_{i},h)=i\pi v_{i}h-\left(  \ln2\right)  /4$ for the toric
code example. Ref.\cite{deng2016exact} gives another construction of RBM representation of the
toric code state, and compared with that our construction here is
significantly simpler.

For the randomly distributed entangled pair states, we have $g(v_{1}%
,v_{2})=\delta_{v_{1}v_{2}}$ for each entangled pair $|00\rangle+|11\rangle$.
The weight function $W(v_{i},h)$ is a special case of the phase gadget
$W_{\theta}$ with $\theta=0$ (the identity gadget).

For a coherent thermal state defined as
\begin{equation}
|\Psi_{ch}\rangle=\sum_{\mathbf{v}}%
%TCIMACRO{\dprod \limits_{\left\langle i,j\right\rangle }}%
%BeginExpansion
{\displaystyle\prod\limits_{\left\langle i,j\right\rangle }}
%EndExpansion
e^{-\beta v_{i}v_{j}/2}|\mathbf{v}\rangle,
\end{equation}
it has the same correlation function as the corresponding thermal state with
$\beta$ denoting the inverse temperature \cite{verstraete2006criticality}. We
consider the binary variables $\mathbf{v}$ defined on a square lattice as
shown in Fig. \ref{fig:examples}, and the state $|\Psi_{ch}\rangle$ then
defines a coherent thermal state for a 2D Ising model which has a phase
transition at $\beta=\beta_{c}$. At this critical point $\beta_{c}$, the state
$|\Psi_{ch}\rangle$ describes a many-body entangled state for a critical
system. For any value of $\beta$, the wave function of the state $|\Psi
_{ch}\rangle$ can be simply represented by a RBM as shown in Fig.
\ref{fig:examples}. We have $g(v_{1},v_{2})=e^{-\beta v_{1}v_{2}/2}$ for each
pair of visible neurons. The correlation is identical to the case that we have
considered for the DBM representation of a fully connect Boltzmann machine
(see the method section of the main text). We just need a single hidden neuron
to generate this correlation with the weight function given in the method
section of the main text where $J$ is replaced by $-\beta/2$.

\subsection{Proof of theorem 1 under approximation representation}

\begin{figure}[ptb]
\centering
\includegraphics[width=1\linewidth]{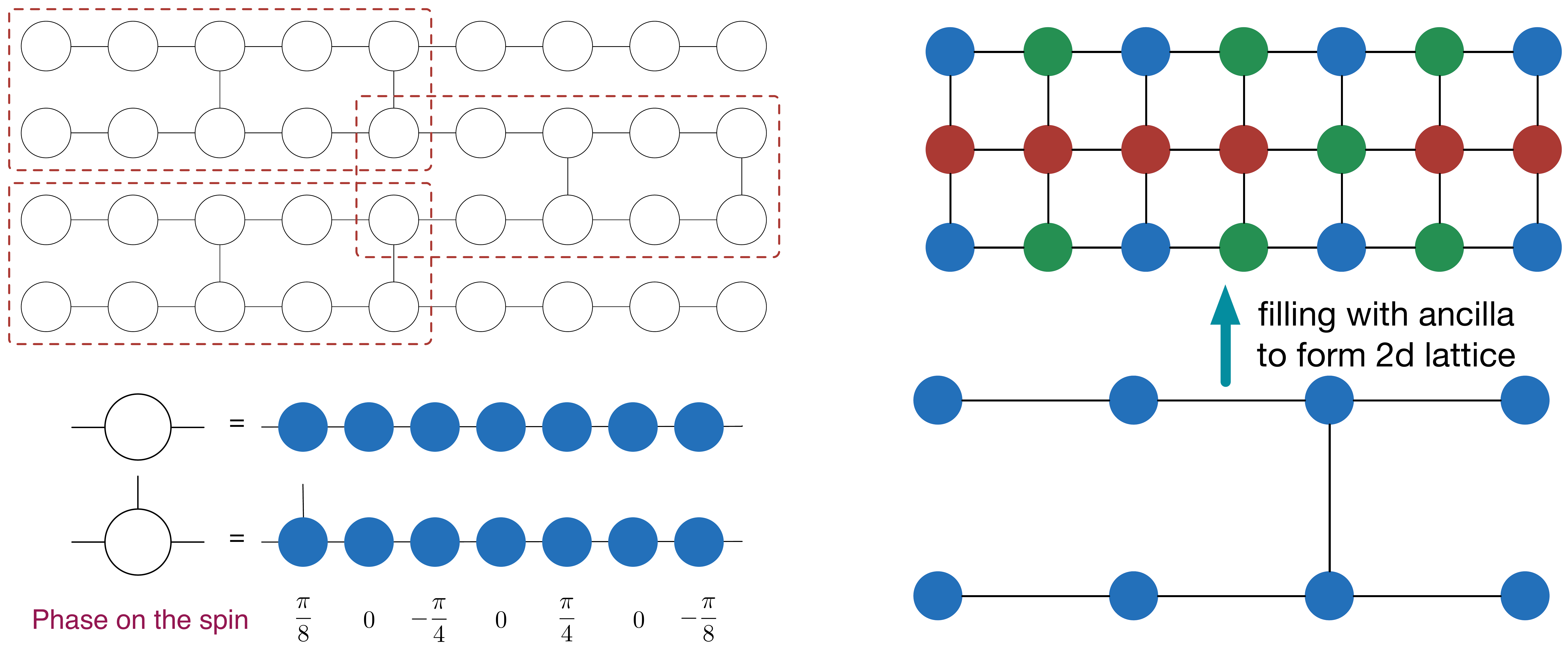}\caption{The state $|\psi_{\rm GWD}\rangle$
used in the proof of theorem 1, which is introduced in Ref. \cite{gao2016quantum} for proof of quantum Supremacy.
To construct this state, we start from a brickwork of white circles \cite{broadbent2009universal} (the left top side), with each white circle
represented by seven blue circles. Each blue circle represents a qubit, and the brickwork of
blue circles can be filled with additional red and green circles (each represent an ancillary qubit) to form a regular 2D square lattice
(shown on the right top side). 
We start with a standard 2D cluster state for the square lattice, and then apply the phase gates $Z(\theta)$ on the blue-circle qubits with the angle 
$\theta$ forming a periodic pattern shown in the left bottom figure and the Hadamard 
gate on each green-circle and blue-circle qubits (no gate on the red-circle qubits).
After this layer of single-qubit unitary operations, we get the state $|\psi_{\rm GWD}\rangle$. }%
\label{fig:qs}%
\end{figure}

Here we introduce a specific state on 2D lattice, denoted as $|\psi
_{\rm GWD}\rangle$ and shown in Fig. \ref{fig:qs}. State $|\psi_{\rm GWD}\rangle$ is
cluster state after one layer translation-invariant single-qubit unitary
transformation. This state was introduced in Ref. \cite{gao2016quantum} for
proof of quantum supremacy. We prove here $|\psi_{\rm GWD}\rangle$ cannot be
efficiently represented by RBMs  under reasonable conjectures in complexity
theory. This no-go theorem holds for both exact and approximate representation.

\subsubsection{Exact representation}

Suppose the coefficient of $|\psi_{\rm GWD}\rangle$ in the computational basis is
$\Psi(\mathbf{v})$ (In term of notation, $|\Psi(\mathbf{v})|^{2}$ corresponds
to $q_{x}$ in Ref. \cite{gao2016quantum}). In Ref. \cite{gao2016quantum}, it
is proved that computing $|\widetilde{\Psi}(\mathbf{v})|^{2}$ is
\textsf{\#P}-hard where $|\widetilde{\Psi}(\mathbf{v})|^{2}$ is an estimation
of $|\Psi(\mathbf{v})|^{2}$ such that
\begin{equation}
\left\vert |\Psi(\mathbf{v})|^{2}-|\widetilde{\Psi}(\mathbf{v})|^{2}%
\right\vert \leq\frac{|\Psi(\mathbf{v})|^{2}}{\mbox{poly}(n)}+\frac{c}{2^{mn}%
}\label{eq:additive_error1}%
\end{equation}
where $0\leq c<1/2$ and the lattice size is $n\times m$. This equation implies
computing $|\widetilde{\Psi}(\mathbf{v})|^{2}$ is also \textsf{\#P}-hard such
that
\begin{equation}
\left\vert |\Psi(\mathbf{v})|^{2}-|\widetilde{\Psi}(\mathbf{v})|^{2}%
\right\vert \leq\frac{c}{2^{mn}}.
\end{equation}

Denote $\mathcal{O}$ as an oracle with the ability to compute the first $mn-1$
digits of $\Psi(\mathbf{v})$. The above statement implies
\begin{equation}
\mathsf{P}^{\#P}\subseteq\mathsf{P}^{\mathcal{O}}.
\end{equation}
If $|\psi_{\rm GWD}\rangle$ can be represented efficiently by a RBM, calculating
its wave function in the computational basis belongs to the complexity class
$\mathsf{P/poly}$ as discussed in the main text. Thus $\mathcal{O}%
\subseteq\mathsf{P/poly}$. Combining these results, we have

\begin{lemma}
RBM cannot represent $|\psi_{\rm GWD}\rangle$ exactly unless
\begin{equation}
\mathsf{P}^{\#P}\subseteq\mathsf{P}^{\mathsf{P/poly}}.
\end{equation}

\end{lemma}

Ref. \cite{babai1990nondeterministic} proves if the containment in the above
equation is true, the polynomial hierarchy will collapse
\cite{arora2009computational}, which is widely believed to be unlikely.

\subsubsection{Approximate representation}

Denote a measurement in the computational basis as a quantum operator
$\mathcal{E}$ and $\widetilde{\Psi}(\mathbf{v})$ as the wave function of a RBM
state $|\psi_{\rm GWD}^{\prime}\rangle$ to approximate $|\psi_{\rm GWD}\rangle$, then
we have
\begin{equation}
\sum_{\mathbf{v}}\left\vert |\Psi(\mathbf{v})|^{2}-|\widetilde{\Psi
}(\mathbf{v})|^{2}\right\vert =2D(\mathcal{E}(|\psi_{\rm GWD}\rangle\langle
\psi_{\rm GWD}|),\mathcal{E}(|\psi_{\rm GWD}^{\prime}\rangle\langle\psi_{\rm GWD}^{\prime
}|))\leq 2D(|\psi_{\rm GWD}\rangle\langle\psi_{\rm GWD}|,|\psi_{\rm GWD}^{\prime}%
\rangle\langle\psi_{\rm GWD}^{\prime}|)\leq\epsilon,
\end{equation}
where $D\left( \rho_1,\rho_2\right)  $ denotes the trace distance, defined as $\mbox{tr}|\rho_1-\rho_2|/2$. The above equation means if the trace distance between $|\psi
_{\rm GWD}\rangle$ and $|\psi_{\rm GWD}^{\prime}\rangle$ is smaller than $\epsilon/2$,
we have $\mathbb{E}_{\mathbf{v}}\left[  \left\vert |\Psi(\mathbf{v}%
)|^{2}-|\widetilde{\Psi}(\mathbf{v})|^{2}\right\vert \right]  \leq
\epsilon/2^{mn}$ where $\mathbb{E}_{\mathbf{v}}[f(\mathbf{v})]$ means the expectation value of $f(\mathbf{v})$ over uniform distribution of $\mathbf{v}$. Using the Markov
inequality, we get
\begin{equation}
\Pr_{\mathbf{v}}\left[  \left\vert |\Psi(\mathbf{v})|^{2}-|\widetilde{\Psi
}(\mathbf{v})|^{2}\right\vert \geq\frac{\epsilon}{2^{mn}\delta}\right]
\leq\delta,
\end{equation}
where $\epsilon/\delta<1/2$ and $\Pr_{\mathbf{v}}[f(\mathbf{v})]$ denotes the probability such that $\mathbf{v}$ satisfies condition $f(\mathbf{v})$ if random variable $\mathbf{v}$ is uniform distributed.
This equation means that for $1-\delta$ fraction of the whole set $\mathbf{v}%
$, we have
\begin{equation}
\left\vert |\Psi(\mathbf{v})|^{2}-|\widetilde{\Psi}(\mathbf{v})|^{2}%
\right\vert \leq\frac{\epsilon}{2^{mn}\delta}.\label{eq:additive_error2}%
\end{equation}
Denote $\mathcal{O}^{\prime}$ as an oracle with the ability to compute the
first $mn-1$ digits of $|\Psi(\mathbf{v})|^{2}$ for such fraction of
$\mathbf{v}$, then $\mathcal{O}^{\prime}\subseteq\mathcal{O}\subseteq
\mathsf{P/poly}$.

In Ref. \cite{gao2016quantum}, we introduced a conjecture that \textsf{\#P}%
-hardness of approximating $\Psi(\mathbf{v})$ to the error given by Eq.
(\ref{eq:additive_error2}) still holds when we lift from the worst-case to the
average-case, that is,

\begin{conjecture}
\label{thm:conjecture} For any $1-\delta$ fraction of instance $\mathbf{v}$,
approximating $|\Psi(\mathbf{v})|^{2}$ by $|\widetilde{\Psi}(\mathbf{v})|^{2}$
up to the error
\[
\left\vert |\Psi(\mathbf{v})|^{2}-|\widetilde{\Psi}(\mathbf{v})|^{2}%
\right\vert \leq\frac{\epsilon}{2^{mn}\delta}%
\]
is still \textsf{\#P}-hard.
\end{conjecture}

Ref. \cite{gao2016quantum} has discussed why this is a reasonable conjecture.
It is related to classical-hardness for simulating distribution from random
quantum circuit and is supported by both quantum chaos theory and extensive
numerical simulations \cite{boixo2016characterizing}. With this conjecture, we
have
\begin{equation}
\mathsf{P}^{\#P}\subseteq\mathsf{P}^{\mathcal{O}^{\prime}}.
\end{equation}
Combining the results above, we have the following theorem

\begin{lemma}
If the conjecture \ref{thm:conjecture} is true, any RBM states cannot
approximate $|\psi_{\rm GWD}\rangle$ with the trace distance smaller than
$\epsilon/2$, otherwise $\mathsf{P}^{\#P}\subseteq\mathsf{P}^{\mathsf{P/poly}}$
and the polynomial hierarchy collapses.
\end{lemma}

\subsection{Efficient tensor network representation for ground states}

\begin{figure}[ptb]
\label{fig:taylor} \centering
\includegraphics[width=1\linewidth]{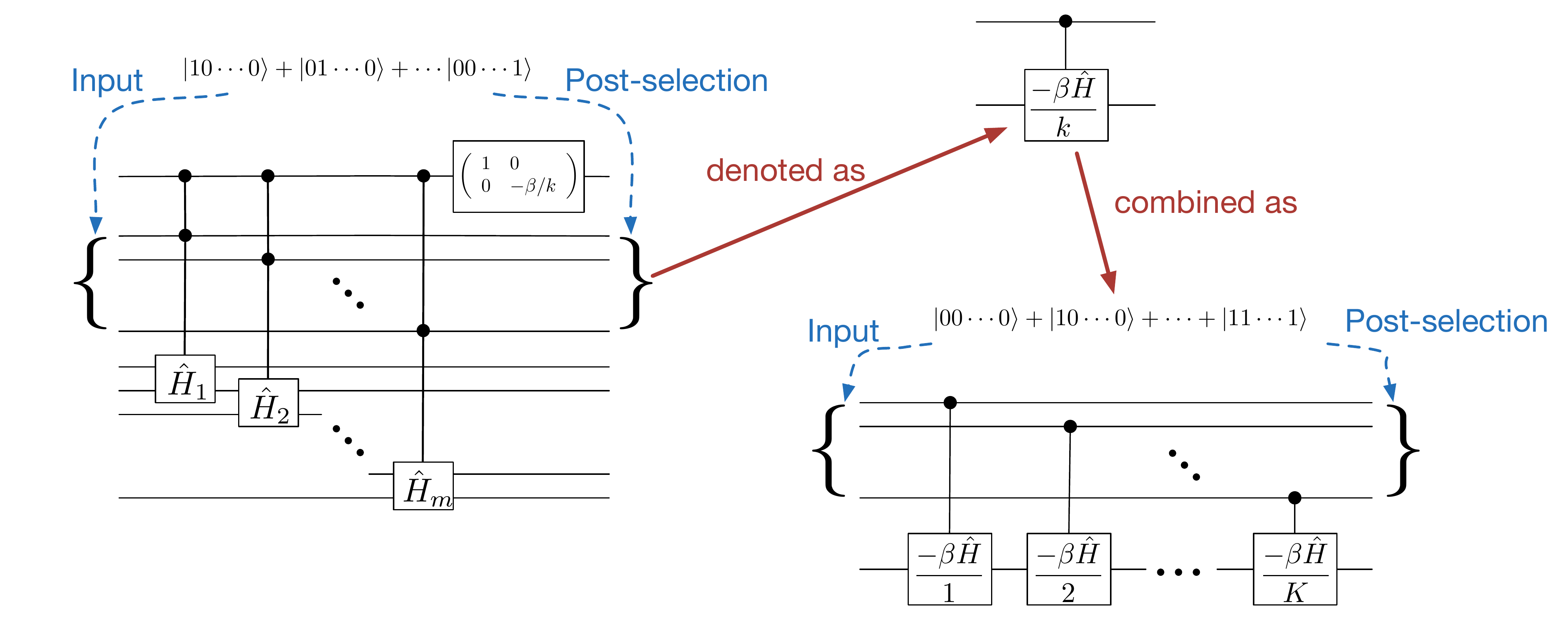}\caption{Construction of pseudo quantum circuit
where each elementary gate represents a linear but non-unitary operation which can be described through 
multiplication of a local tensor. The left diagram represents construction of the pseudo-gate
controlled-$(-\beta\hat{H}/k)$, which acts as a basic building block for construction of the pseudo-gate 
$e^{-\beta\hat{H}}$ shown on the lower right side through the Taylor series expansion.}%
\label{fig:taylor}%
\end{figure}

In proof of theorem 4, we mention that we have developed a method using Taylor
series expansion to efficiently construct ground states of any $k$-local
Hamiltonians with tensor network (and thus DBM network as well due to theorem
3). Compared with the previous construction method \cite{schuch2007computational}, this approach allows an
exponential improvement in the precision of the representation. We use pseudo
quantum circuit to present our construction as shown in Fig. \ref{fig:taylor}.
Pseudo quantum circuit is similar to conventional quantum circuit except that
the pseudo-gate is not required to be unitary. Each pseudo-gate still represents 
a linear transformation through matrix multiplication, so it can be easily constructed through
a local tensor. The pseudo quantum circuit then just represents a tensor network.

Suppose the number of interaction terms in the Hamiltonian is $m$, i.e.,
$\hat{H}\equiv\sum_{i=1}^{m}\hat{H}_{i}$, where each $\hat{H}_{i}$ involves at
most $k$-body interactions ($k$ is typically a small finite constant). We
simulate the operator $\hat{H}$ by first generating a state $\sum_{i=1}%
^{m}|i\rangle$, where $|i\rangle\equiv|0_{1}0_{2}\cdots1_{i}\cdots0_{m}%
\rangle$, i.e., only the $i$-th bit is $1$. Applying the operation $\hat
{H}_{i}$ controlled by the $i$-th qubit as shown in Fig. \ref{fig:taylor} and
post-selecting the control bits in the state $\sum_{i=1}^{m}|i\rangle$ (note
that postselection can be easily represented in a tensor network), we get
\begin{equation}
\sum_{i=1}^{m}|i\rangle\xrightarrow{\mbox{controlled-}\hat H_i}\sum_{i=1}%
^{m}|i\rangle\hat{H}_{i}\xrightarrow{\mbox{post-selection}}\sum_{i=1}^{m}%
\hat{H}_{i}.
\end{equation}
This requires $O(m)$ pseudo quantum gates as shown in Fig. \ref{fig:taylor}.
All of the above operations are controlled by an additional qubit (see Fig.
\ref{fig:taylor}). We then apply a non-unitary matrix $\mbox{diag}(1,-\beta
/k)$ on this control qubit, and construct a pseudo-gate controlled-$(-\beta
\hat{H}/k)$. Note that this gate requires $O(m)$ elementary pseudo quantum
gates for its construction and we use the pseudo-gate controlled-$(-\beta
\hat{H}/k)$ as a building block for the next step. In the next step, we first
generate a superposition state $\sum_{k=0}^{K}|k\rangle$, where $K$
corresponds to the truncation number in Taylor series expansion and
$|k\rangle\equiv|1_{1}1_{2}\cdots1_{k}0_{k+1}\cdots0_{K}\rangle$, i.e. only
the first $k$ bits are in state $|1\rangle$. Applying the circuit shown in
Fig. \ref{fig:taylor} and post-selecting the output state of the control
qubits in $\sum_{k=0}^{K}|k\rangle$, we get
\begin{equation}
\sum_{k=0}^{K}|k\rangle\xrightarrow{\mbox{controlled-}(-\beta \hat H/k)}\sum
_{k=0}^{K}|k\rangle\frac{(-\beta\hat{H})^{k}}{k!}%
\xrightarrow{\mbox{post-selection}}\sum_{k=0}^{K}\frac{(-\beta\hat{H})^{k}%
}{k!}%
\end{equation}
which is the Taylor expansion of $e^{-\beta\hat{H}}$ truncated to the $K$-th
order. Note that this construction of $e^{-\beta\hat{H}}$ requires $O(Km)$
elementary pseudo quantum gates.

We apply the above operator $e^{-\beta\hat{H}}$ on the state
\begin{equation}
\sum_{i=0}^{2^{n}-1}|i\rangle|i\rangle=\sum_{i=0}^{2^{n}-1}|\psi_{i}\rangle
|\psi_{i}^{\ast}\rangle
\end{equation}
where $n$ represents the total number of qubits in the Hamiltonian $\hat{H}$,
$|\psi_{i}\rangle$ denotes the $i$-th eigenstate of $\hat{H}$, and $|\psi
_{i}^{\ast}\rangle$ is the complex conjugate of $|\psi_{i}\rangle$ in the
computational basis. After tracing out the register storing $|\psi_{i}^{\ast
}\rangle$ and dropping the unimportant normalization factor, we get the state%
\begin{equation}
|\psi_{0}\rangle\langle\psi_{0}|+O\left(  2^{n}e^{-\beta\Delta}+\frac
{2^{n}(\beta\Vert\hat{H}\Vert)^{K}}{K!}\right)
\end{equation}
for the first register. In the equation above, the first term represents our
targeted ground state $|\psi_{0}\rangle$, and there are two error terms: the
first term comes from contribution of all the other eigenstates, shrunk by the
imaginary time evolution factor $e^{-\beta\Delta}$ with $\Delta$ denoting the
energy gap; and the second term comes from the truncation error in Taylor
series expansion. Suppose we require the total representation error is bounded
by a small constant $\epsilon$, then we need
\begin{align}
O\left(  2^{n}e^{-\beta\Delta}\right)   &  \leq\frac{\epsilon}{2},\\
O\left(  \frac{2^{n}(\beta\Vert\hat{H}\Vert)^{K}}{K!}\right)   &  \leq
\frac{\epsilon}{2}.
\end{align}
With an optimal choice of the parameter $\beta=O((n+\log(1/\epsilon))/\Delta
)$, we need $K=O(\beta\Vert\hat{H}\Vert)$ to satisfy these inequalities. As
$\Vert\hat{H}\Vert=O(m)$, we have $K=O((n+\log(1/\epsilon))m/\Delta)$. So the
total number of elementary tensors we need to represent the ground state
$|\psi_{0}\rangle$ of the Hamiltonian $\hat{H}$ is given by $O(Km)$, which is
\begin{equation}
O\left(  \frac{1}{\Delta}\left(  n+\log\frac{1}{\epsilon}\right)
m^{2}\right)  .
\end{equation}
Each elementary tensor has a constant bond dimension $D$ and a typically small
coordination number $d$. Combining with theorem 3 in the main text, we find
the total number of neurons in the DBM\ to represent the ground state of the
Hamiltonian $\hat{H}$ is also given by the above equation, which is the
statement of theorem 4.

\end{document}